\documentclass[annual]{acmsiggraph}

\TOGonlineid{0000}

\TOGvolume{30}
\TOGnumber{4}

\TOGarticleDOI{1964921.1964951}

\TOGprojectURL{}
\TOGvideoURL{}
\TOGdataURL{}
\TOGcodeURL{}

\title{An analytic BRDF for materials with spherical Lambertian scatterers}

\author{ Eugene d'Eon \\
NVIDIA }

\pdfauthor{ }

\keywords{BRDF, porous, H-function, Lambertian, sphere, granular media}

\usepackage{smrdefaults}

\usepackage{color}

\usepackage{multirow}

\usepackage{rotating}
\newcommand{\vect}[1]{\vec{#1}}

\newcommand{\n}{\vect{\mathbf{n}}}

\newcommand{\dir}{\vect{\omega}}

\def\rm{R^{-}}

\usepackage{paralist}
\usepackage{subfigure}
\usepackage{rotating}
\usepackage{amsgen}

\usepackage{alltt}
\usepackage{cuted}

\begin{document}


 \teaser{
   \centering
   \includegraphics[width=\linewidth]{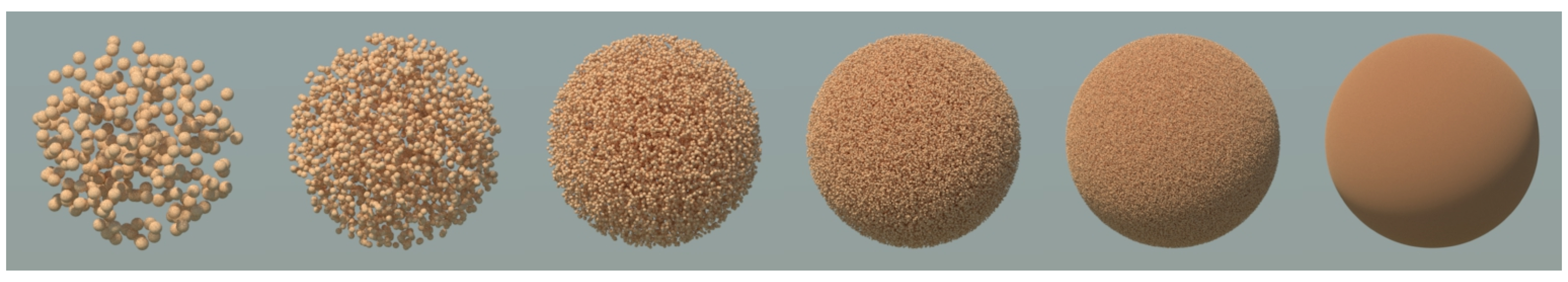}
    \caption{We derive a new analytic BRDF (right) for porous materials where scattering and absorption is well approximated by spherical Lambertian particles.  Here we show a sequence of stochastic microgeometries consisting of independent random placement of opaque Lambertian spheres occupying $7\%$ of the volume.  As the scale of the microgeometry is varied inside the fixed spherical domain, the appearance approaches our BRDF where no spatial variability is resolved across pixels.}
    \label{fig-teaser}
 }


\maketitle


\begin{abstract}
  \large We present a new analytic BRDF for porous materials comprised of spherical Lambertian scatterers.  The BRDF has a single parameter: the albedo of the Lambertian particles.  The resulting appearance exhibits strong back scattering and saturation effects that height-field-based models such as Oren-Nayar cannot reproduce.
\end{abstract}




\keywordlist






\section{Introduction}

\large

The \emph{bidirectional reflectance distribution function} (BRDF) is a fundamental building block in computer graphics and other fields.  BRDF measurements have shown that real world materials exhibit a wide range of reflectance behaviours~\cite{matusik2003efficient,dupuy2018adaptive}.  While measured data can be used directly, it is bulky and difficult to edit.  In contrast, parametric BRDFs are compact and permit artist control, but no one parametric BRDF spans the full breadth of real-world appearances.  It is therefore important to define a small set of flexible analytic parametric BRDFs that cover a wide range of materials.

The most popular parametric BRDFs derive from Smith's \shortcite{smith1967geometrical} geometrical and statistical treatment of random height fields \cite{torrance67,cook82,blinn82,vanginneken98,stam01,walter07,heitz2014understanding}.  These models support a variety of surface statistics \cite{ribardiere2017std}, anisotropy \cite{heitz2014understanding}, importance sampling \cite{heitz2014importance} and multiple scattering with specular, diffuse, or mixed microfacets \cite{heitz2015implementing,heitz2016multiple,meneveaux2017rendering}.  These BRDFs have been highly successful and cover a wide range of materials, but the assumption of a height field is not always valid~\cite{dupuy2016additional}: these BDRFs cannot model \emph{porous/granular/volumetric/particulate} microgeometry, such as the example shown in \autoref{fig-teaser}, or the electron photomicrograph images of materials like carbon soot \cite{shkuratov2008laboratory}.

\emph{Volumetric} BRDFs can be derived by evaluating a plane-parallel projection of scalar radiative transfer in slab and halfspace geometries, which statistically accounts for the interaction of light with randomly distributed absorbing and scattering particles in a volume \cite{vandehulst80,yanovitskij1997light}.  Multiple volumetric slabs can be interleaved between random height fields to form fully general layer stacks \cite{stam01} that can be stochastically evaluated in a ``position-free'' way \cite{guo2018position} or numerically pretabulated using adding/doubling or related methods \cite{stam01,jakob2014comprehensive,zeltner2018layer,belcour2018efficient}.  A great variety of parametric layered materials can be evaluated using these methods, but they can significantly exceed the complexity, cost and memory requirements of simpler analytical BRDFs.  One exception is the half space with isotropic scattering, which has a known semi-analytic BRDF \cite{chandrasekhar60,premoze2002analytic} (Chandrasekhar's BRDF) as well as an accurate fully-analytic  approximation \cite{keller2020advances}.  

Chandrasekhar's BRDF is the unique volumetric BRDF that can be applied as efficiently and broadly as other analytic height-field BRDFs.  It has inspired BRDFs for lunar regoliths \cite{hapke81} and has been extended to support Fresnel reflection at the boundary~\cite{wolff1998improved,williams06}.  A stochastic geometry that is consistent with Chandrasekhar's BRDF (under geometrical-optics) is a sparse suspension of mirror sphere particles in a purely absorbing substrate.  In this paper we derive what is the Oren-Nayar \shortcite{oren1994generalization} equivalent of this BRDF, by making the scatterers Lambertian.  By the equivalence principle \cite{vandehulst80}, this is equivalent to a void substrate with absorbing Lambertian spherical particles.  The result is a new BRDF for dusty/porous materials (\autoref{fig-teaser}) that differs from alternative diffusive analytic BRDFs.

\section{Lambertian sphere phase function}
  The far-field geometrical optics phase function for a smooth, white Lambertian sphere is~\cite{schoenberg1929theoretische,blinn82,rushmeier1995input,porco2008simulations}
    \begin{equation}\label{eq:phase:lambertsphere}
      p(\mu) = \frac{2 \left(\sqrt{1-\mu^2}-\mu \cos ^{-1}(\mu)\right)}{3 \pi ^2}
    \end{equation}
    where $\mu = \cos \theta$.
    The mean cosine is $g = -4/9 \approx -0.444444$.  For the Legendre expansion of this phase function 
    \begin{equation}
      p(\mu) =  \frac{1}{4\pi} \sum_{k=0}^\infty A_k P_k(\mu),
    \end{equation}
    where expansion coefficients are defined as
    \begin{equation}
      A_k =  2 \pi (2 k + 1) \int_{-1}^1 p(\mu) P_k(\mu) d\mu,
    \end{equation}
    we find the first few expansion coefficients
    \begin{equation}\label{eq:As}
      A_0 = 1, \quad  A_1 = -\frac{4}{3}, \quad A_2 = \frac{5}{16}, \quad A_3 = 0, \quad A_4 = \frac{1}{64}, \quad A_5 = 0, \quad A_6 = \frac{13}{4096}.
    \end{equation}
    \begin{figure}
      \centering
      \includegraphics[width=.5\linewidth]{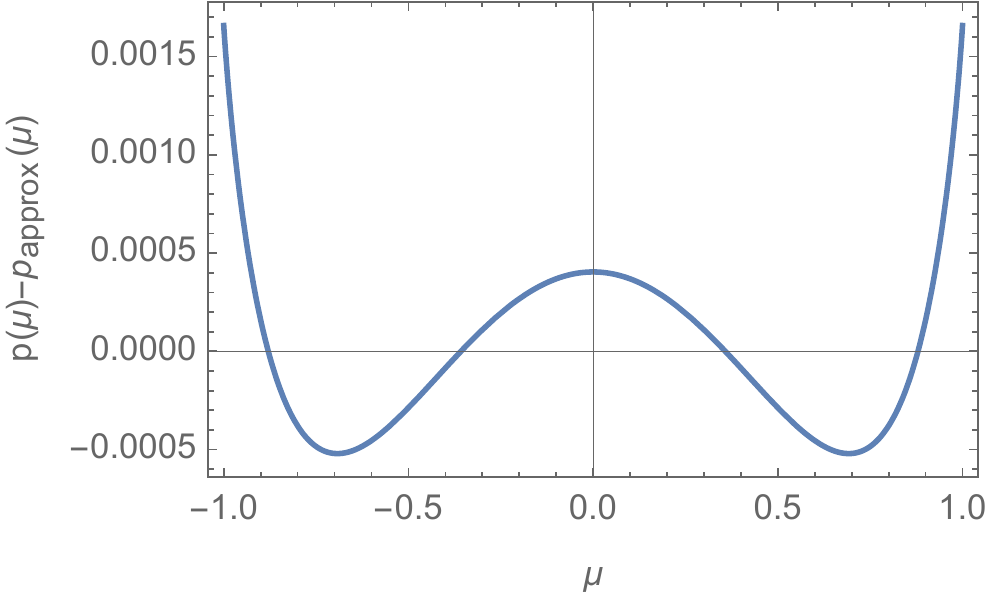}
      \caption{A 3-term Legendre expansion of the phase function of the Lambertian sphere particle is accurate to within $0.2\%$.}
      \label{fig-null-sphere} 
    \end{figure}
    Observing that the majority of the phase function is represented by the first three terms in the expansion (\autoref{fig-null-sphere}), we approximate the phase function using only these first three terms, $A_k \approx 0, k \ge 3$.  This permits us to apply an exact derivation for half space BRDFs \cite{horak61}, which we use to derive a practical fully-analytic approximation.

    \subsubsection{Importance Sampling}
    For ground-truth validation of our approximate model, we use Monte Carlo simulation of particle transport in a homogeneous half space with the Lambert-sphere phase function (\autoref{eq:phase:lambertsphere}).  To the best of our knowledge, there is no published procedure for importance-sampling this phase function, and we propose two such procedures here.  For an exact result, we note that deflection cosines can be randomly sampled with
    \begin{equation}
      \mu(\xi_1,\xi_2,\xi_3) = \sqrt{(1-\xi_1)(1-\xi_2)} \sin \left( 2 \pi \xi_3 \right) -\sqrt{\xi_1 \xi_2}
    \end{equation}
    where $\xi_1, \xi_2, \xi_3$ are three independent random numbers drawn uniformly from $[0,1)$.
    Alternatively, with a single uniform random real $\xi$, $\mu$ can be sampled using an approximate inverse CDF,
    \begin{equation}
      \mu(\xi) \approx 1-2 \left(1-\xi ^{0.0401885 \xi +1.01938}\right)^{0.397225},
    \end{equation}
    which has a maximum absolute error of $|\mu - \mu_{exact}| < 0.0005$.

  \section{BRDF derivation}

    Horak and Chandrasekhar~\shortcite{horak61} derive the exact BRDF of a half space with a general three-term phase function,
    \begin{equation}
      c \, p(\cos \theta) = \varpi_0 + \varpi_1 P_1(\cos \theta) + \varpi_2 P_2(\cos \theta)
    \end{equation}
    where $P_1$ and $P_2$ are Legendre polynomials.  In classic radiative transfer notation, the \emph{single-scattering albedo} $c$ (which here is the diffuse albedo of the spherical particles) is folded into the phase function and so, in the case of the 3-term truncation given by coefficients in \autoref{eq:As}, we have
    \begin{equation}\label{eq:varpi}
      \varpi_0 = c, \quad \varpi_1 = \frac{-4 c}{3}, \quad \varpi_2 = \frac{5 c}{16}.
    \end{equation}
    The BRDF of a three-term half space is given exactly as a sum of three Fourier modes 
    \begin{equation}
      f_r(\dir_i,\dir_o) = f^{(0)}(\mu_i,\mu_o) + f^{(1)}(\mu_i,\mu_o) \cos(\phi) + f^{(2)}(\mu_i,\mu_o) \cos(2 \phi).
    \end{equation}
    The functions $f^{(i)}(\mu_i,\mu_o)$ are cone-to-cone transfer functions and are closely related to transfer matrices used in adding/doubling and related numerical methods \cite{vandehulst80,jakob2014comprehensive}.  For light arriving from a direction with cosine $\mu_i$, the integrated radiance leaving the material along the cone with cosine $\mu_o$ is $f^{(0)}(\mu_i,\mu_o)$, and the higher order terms give the discrete cosine series that determine the variation of the outgoing radiance within that cone, parametrized by relative azimuth $\phi$.

    The functions $f^{(i)}(\mu_i,\mu_o)$ include all orders of scattering and are complex expressions in the case of a three-term phase function.  We would like to take advantage of the known simple analytic expression for the single-scattering component of the BRDF \cite{chandrasekhar60,hanrahan93}
    \begin{equation}\label{eq:fsingle}
      f_1(\dir_i,\dir_o) = c \frac{p(-\dir_i \cdot \dir_o )}{\mu_i + \mu_o}.
    \end{equation}
    To exploit this result, and to further ensure that single-scattering is represented exactly, we will represent our final BRDF as the sum of single-scattering and multiple-scattering terms
    \begin{equation}\label{eq:BRDF}
      f_r(\dir_i,\dir_o) = f_1(\dir_i,\dir_o) + f_m(\dir_i,\dir_o)
    \end{equation}
    where the single-scattering portion is computing using \autoref{eq:fsingle}.
    We will therefore use the three-term expansion of the phase function only for solving for the multiple-scattering components 
    \begin{equation}
      f_m(\dir_i,\dir_o) = f_m^{(0)}(\mu_i,\mu_o) + f_m^{(1)}(\mu_i,\mu_o) \cos(\phi) + f_m^{(2)}(\mu_i,\mu_o) \cos(2 \phi).
    \end{equation}
    These functions can be determined from the general derivation \cite{horak61} once the corresponding $H$ functions and constants are solved for.

    \subsection{The $H$ functions}

    The BRDF for a half space with a three-term phase function can be derived using invariance principles.  This derivation leads to three pseudo problems with three corresponding $H$ functions.  The characteristic functions $\Psi^{(i)}(\mu)$ for these $H$ functions follow from inserting \autoref{eq:varpi} into the general solution \cite[p.55]{horak61}, and we find
    \begin{align}
      \Psi^{(0)}(\mu) &= \frac{1}{384} c \left(-15 (c-1) (4 c+9) \mu ^4+(c (20 c+281)-346) \mu ^2+207\right),\\
      \Psi^{(1)}(\mu) &= -\frac{1}{192} c \left(\mu ^2-1\right) \left(5 (4 c+9) \mu ^2-64\right),\\
      \Psi^{(2)}(\mu) &= \frac{15}{256} c \left(\mu ^2-1\right)^2.
    \end{align}
    We can then numerically evaluate the $H$ functions using the Fok/Chandrasekhar equation \cite{fock1944some,krein1962integral} 
    \begin{equation}\label{eq:H}
      H^{(i)}(\mu) = \exp \left( -\frac{\mu}{\pi} \int_0^\infty \frac{1}{1+\mu^2 t^2} \log K^{(i)}(t) dt \right),
    \end{equation}
    where the functions $K^{(i)}(t)$ are given by \cite{krein1962integral}
    \begin{equation}
      K^{(i)}(t) = 1 - \int_1^\infty \left( \frac{1}{s-i t} + \frac{1}{s+i t}\right) \frac{\Psi^{(i)}\left(\frac{1}{s}\right)}{s} ds.
    \end{equation}
    Working these out, we find
    \begin{align}
      K^{(0)}(t) &= 1-\frac{c \left((256 c-301) t^3+\left((346-c (20 c+281)) t^2-15 (c-1) (4 c+9)+207
      t^4\right) \tan ^{-1}(t)+15 (c-1) (4 c+9) t\right)}{192 t^5}, \label{eq:K0} \\
      K^{(1)}(t) &= 1-\frac{c \left((40 c+282) t^3-3 \left(t^2+1\right) \left(20 c+64 t^2+45\right) \tan
      ^{-1}(t)+15 (4 c+9) t\right)}{288 t^5}, \label{eq:K1} \\
      K^{(2)}(t) &= 1-\frac{5 c \left(3 \left(t^2+1\right)^2 \tan ^{-1}(t)-t \left(5 t^2+3\right)\right)}{128
      t^5}.
    \end{align}
    We will these with \autoref{eq:H} to numerically evaluate the $H$ functions and form more efficient analytic approximations suitable to direct use in rendering.  Alternatively, the $H$ functions can also be evaluated using quadrature methods~\cite{chandrasekhar60,horak61}.

  \subsection{Second-order Fourier mode}

\begin{figure}
      \centering
      \includegraphics[width=\linewidth]{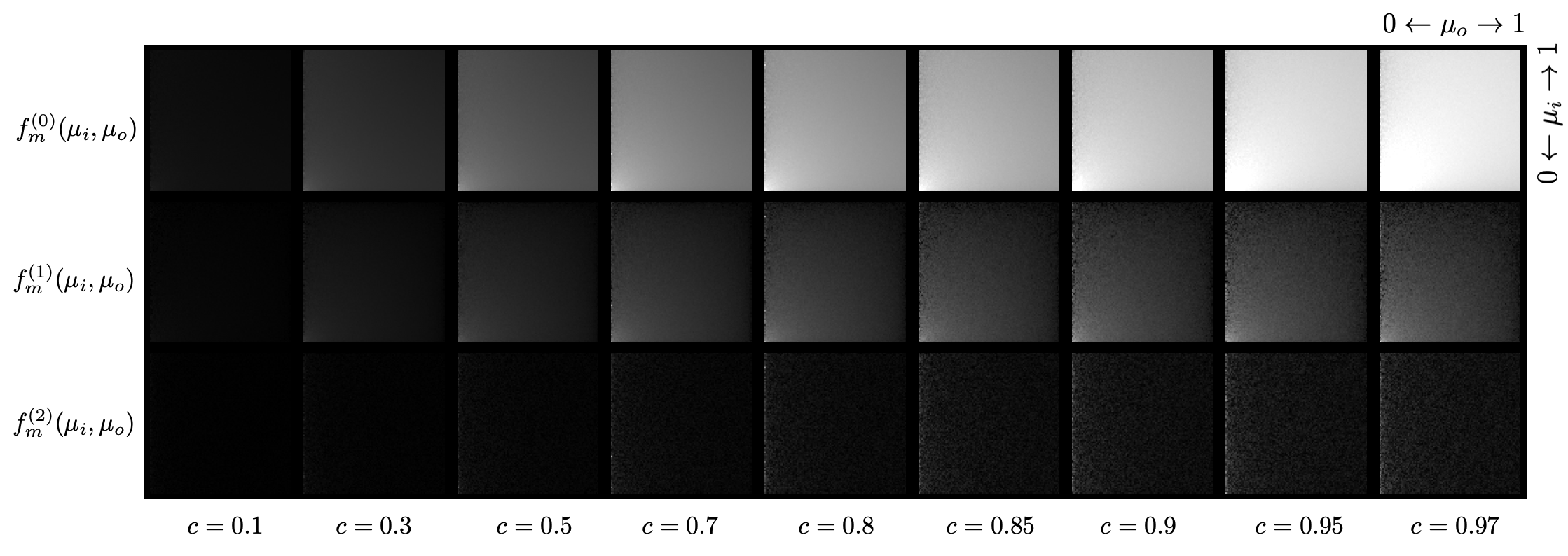}
      \caption{Using Monte Carlo reference, we observe comparatively weak signal in the second-order mode of the multiple-scattering portion of the BRDF, $f^{(2)}(\mu_i,\mu_o)$ (bottom row).}
      \label{fig-f0m-grid} 
    \end{figure}

    For the second-order Fourier mode $f^{(2)}(\mu_i,\mu_o)$, we observe (\autoref{fig-f0m-grid}), using MC reference, that the multiple-scattering component $f_m^{(2)}(\mu_i,\mu_o)$ is very weak when compared to the total energy (and even just the multiply-scattered energy) in the BRDF.  This happens because the already low-frequency phase function is convolved into a nearly linear-cosine shape after two or more collisions.  We exploit this properly to simplify our analytic BRDF by simply setting this term to 0,
    \begin{equation}
      f_m^{(2)}(\mu_i,\mu_o) \cos(2 \phi) \approx 0.
    \end{equation} 

  \subsection{First-order Fourier mode}

  The first-order mode of the BRDF is \cite[Eq.(43)]{horak61}
  \begin{equation}\label{eq:f1}
    f^{(1)}(\mu_i,\mu_o) = \frac{c H^{(1)}\left(\mu _i\right) H^{(1)}\left(\mu _o\right) }{6 \pi \left(\mu _i+\mu _o\right)} \sqrt{\left(1-\mu _i^2\right)
    \left(1-\mu _o^2\right)} \left(1 + \left(l^2+\frac{45 m}{64}\right) \mu _i  \mu _o+l
    \left(\mu _i+\mu _o\right)\right),
  \end{equation}
  requiring determination of two constants $\{l,m\}$, and the $H$ function.

  \subsubsection{$H^{(1)}$ approximation}
  We used \autoref{eq:H} and \autoref{eq:K1} to fit an approximation for $H^{(1)}(\mu)$.  We found that the separable approximation
  \begin{equation}
    H^{(1)}(\mu) \approx H^{(1)}(1) H_{c=1}^{(1)}(\mu)
  \end{equation}
  with
  \begin{equation}
    H_{c=1}^{(1)}(\mu) \approx e^{-0.0894878 \mu ^{-1.12831 \mu ^3+1.85728 \mu ^2-1.07879 \mu +0.459442}}
  \end{equation}
  and
  \begin{equation}
    H^{(1)}(1) \approx e^{0.0242851 c^2-0.144839 c}
  \end{equation}
  was accurate to within $0.5\%$ (relative error).

  \subsubsection{Constants}

\begin{figure}
    \centering
    \includegraphics[width=0.7\linewidth]{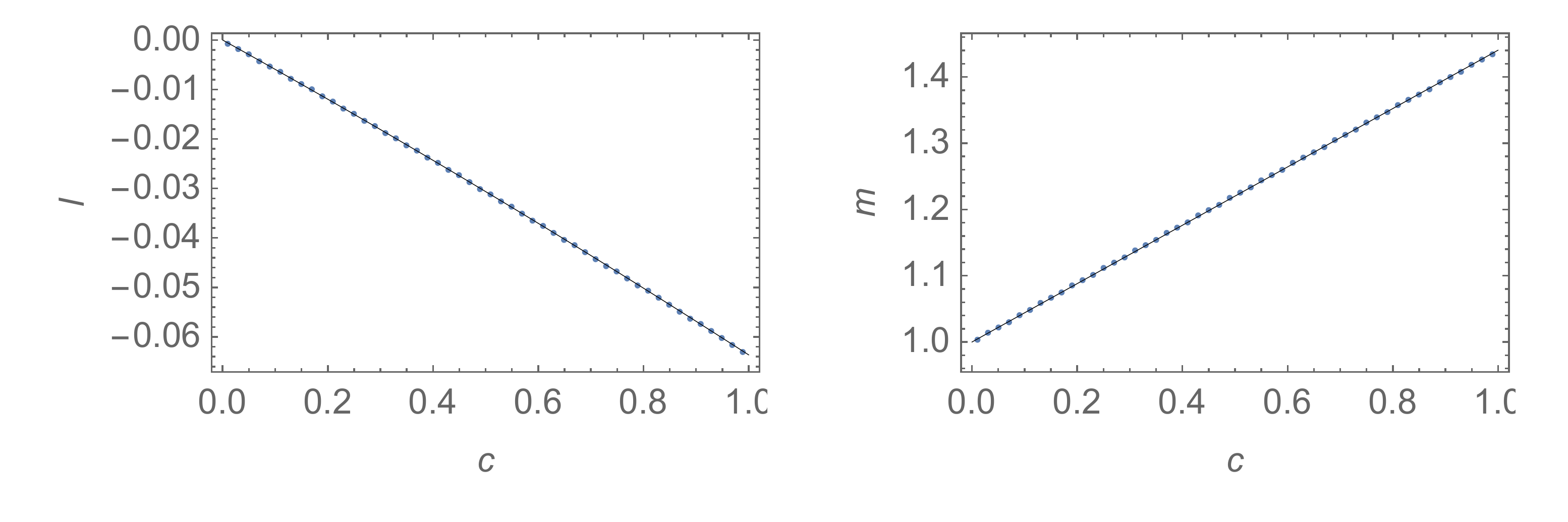}
    \caption{The constants $l$ and $m$ appearing in the first-order mode of our BRDF are well approximated by simple approximations (continuous).}
    \label{fig-val-lm} 
  \end{figure}

  Two constants $\{l,m\}$ appear in the first-order mode, and these follow from moments of the $H$ function \cite[p. 56]{horak61}.  Using numerical evaluation of the $H$ function moments we found the following approximations to be very accurate (\autoref{fig-val-lm}),
  \begin{align}
    l &\approx -0.00473696 c^2-0.0589037 c, \\
    m &\approx 0.44038 c+1. 
  \end{align}

  \subsubsection{Single-scattering}

  \autoref{eq:f1} contains all orders of scattering.  In order to use the single-scattering result exactly, we need to subtract out the approximate single-scattering from $f^{(1)}(\mu_i,\mu_o)$ using the three-term phase function appoximation,
      \begin{equation}
        f_m^{(1)}(\mu_i,\mu_o) = f^{(1)}(\mu_i,\mu_o) - \frac{c \left(45 \mu _i \mu _o+64\right) \sqrt{\left(\mu _i^2-1\right) \left(\mu
        _o^2-1\right)}}{384 \pi  \left(\mu _i+\mu _o\right)}.
      \end{equation}

  \subsubsection{Fast Approximation}

\begin{figure}
    \centering
    \includegraphics[clip, trim=14.0cm 12.0cm 15.0cm 12.0cm, width=\linewidth]{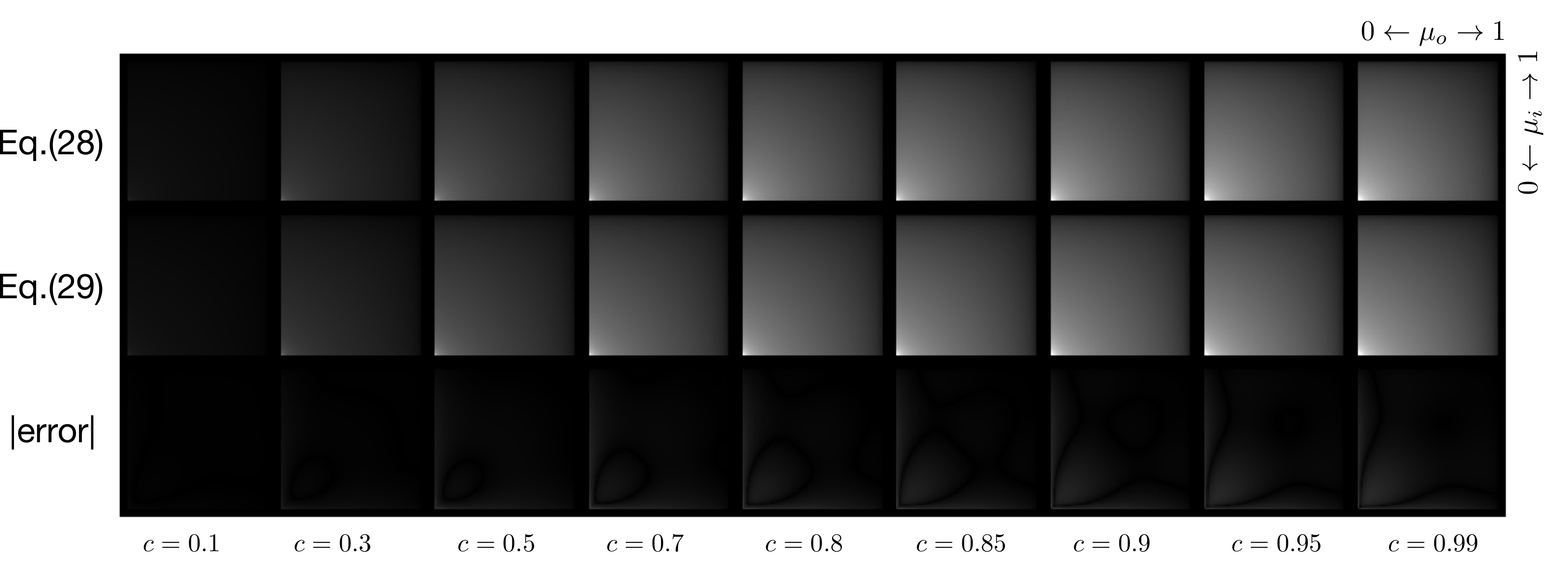}
    \caption{Error analysis of the fast approximation for the first-order transfer matrix $f_m^{(1)}(\mu_i,\mu_o)$.}
    \label{fig-f1m-fast-compare} 
  \end{figure}

  When more efficiency is desired, a less accurate approximation, found using TuringBot symbolic regressions software, is
  \begin{equation}
    f_m^{(1)}(\mu_i,\mu_o) \approx -\frac{0.0117 c \tan ^{-1}(c) \sqrt{\left(1-\mu _i^2\right) \left(1-\mu _o^2\right)}
   \sqrt{\tanh \left(\mu _i+\mu _i \mu _o+\mu _o\right)}}{\mu _i+\mu _o}.
  \end{equation}
  The accuracy of this reciprocal approximation is compared in \autoref{fig-f1m-fast-compare}.

  \subsection{Zeroth-order Fourier mode}

    We use the exact solution in~\cite{horak61} to write $f^{(0)}(\mu_i,\mu_o)$ as
    \begin{equation}\label{eq:f0}
      f^{(0)}(\mu_i,\mu_o) = \frac{1}{2\pi} \frac{H^{(0)}(\mu_i) H^{(0)}(\mu_o)}{\mu_i + \mu_o} \left( A+B \left(\mu _i+\mu _o\right)+ C \mu _i \mu _o+ D \mu _i \mu _o \left(\mu _i+\mu
      _o\right)+ E \mu _i^2 \mu _o^2+F \left(\mu _i^2+\mu _o^2\right)  \right).
    \end{equation}

    \subsubsection{$H^{(0)}$ approximation}
    Evaluation of $f^{(0)}(\mu_i,\mu_o)$ requires numerically integrating $H^{(0)}(\mu)$, using \autoref{eq:H} and \autoref{eq:K0}.  To avoid this cost, we derived an approximate form inspired by Hapke~\shortcite{hapke81},
    \begin{equation}
      H^{(0)}(\mu) \approx \frac{1 + a \mu^d}{1 + \frac{a \mu^d}{H^{(0)}(\infty)}},
    \end{equation}
    where the value at infinity is \cite{ivanov1994resolvent}
    \begin{equation}
      H^{(0)}(\infty) = \frac{1}{\sqrt{K^{(0)}(0)}} = \frac{1}{\sqrt{1-2 \left(-\frac{c^3}{72}+\frac{59 c^2}{288}+\frac{89 c}{288}\right)}} = \frac{12}{\sqrt{(c-16) (c-1) (4 c+9)}}.
    \end{equation}
   We then used numerical-fitting methods to solve for constants $a$ and $d$,
    \begin{equation*}
      a = \frac{1.50112 s^{6.05435}+8.21644}{4.17593\, -1.21222 s}, \quad d = \frac{7.7731\, -0.565811 s^{0.961546}}{8.65912\, -0.159974 s^7}, \quad s = \sqrt{1-c}.
    \end{equation*}
    We found this to have a relative error of less than $1\%$ in the range $\{\mu,c\} \in [0,1]$.

    \subsubsection{Constants}

\begin{figure}
      \centering
      \includegraphics[width=.24\linewidth]{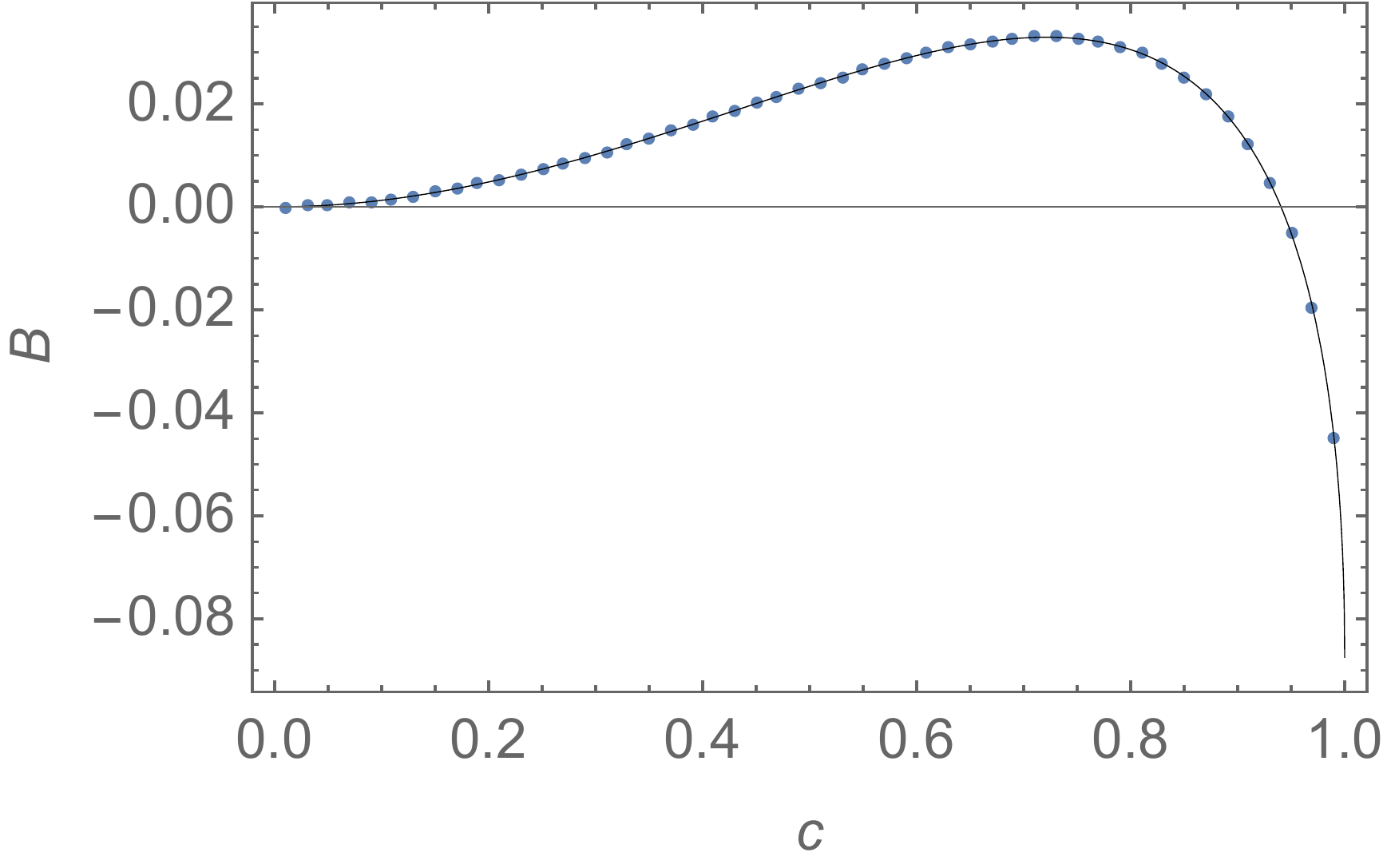}
      \includegraphics[width=.24\linewidth]{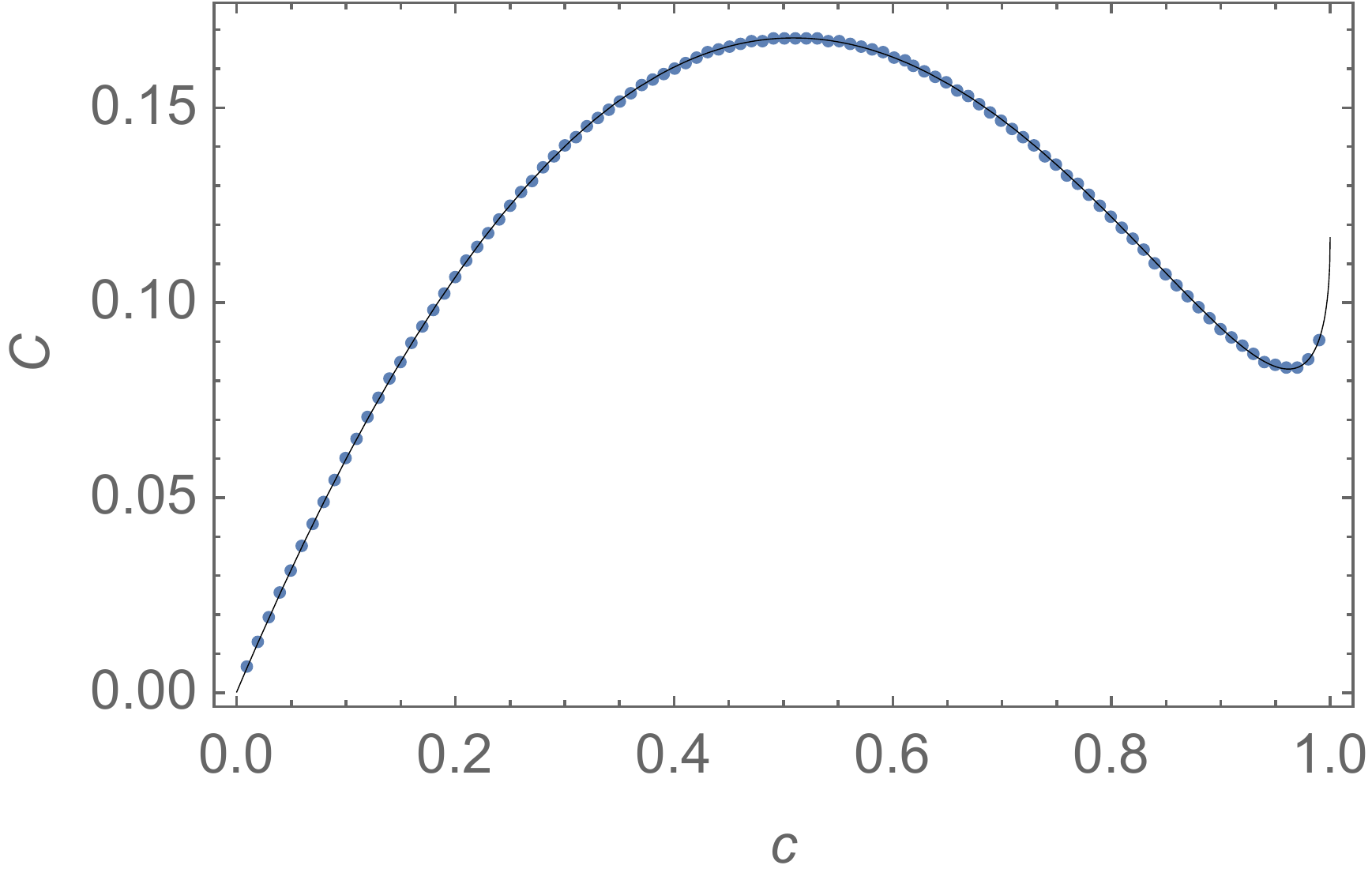}
      \includegraphics[width=.24\linewidth]{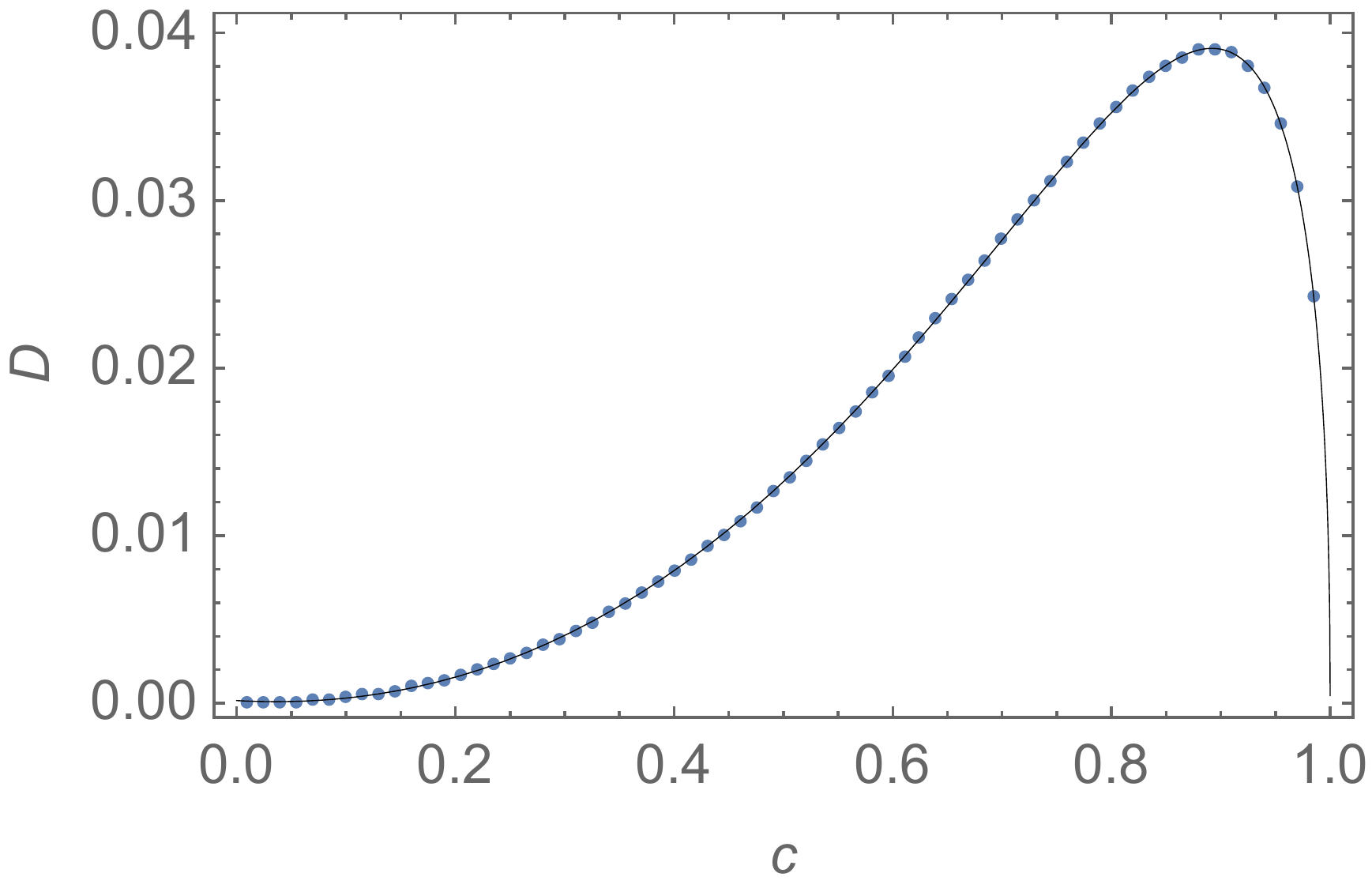}
      \includegraphics[width=.24\linewidth]{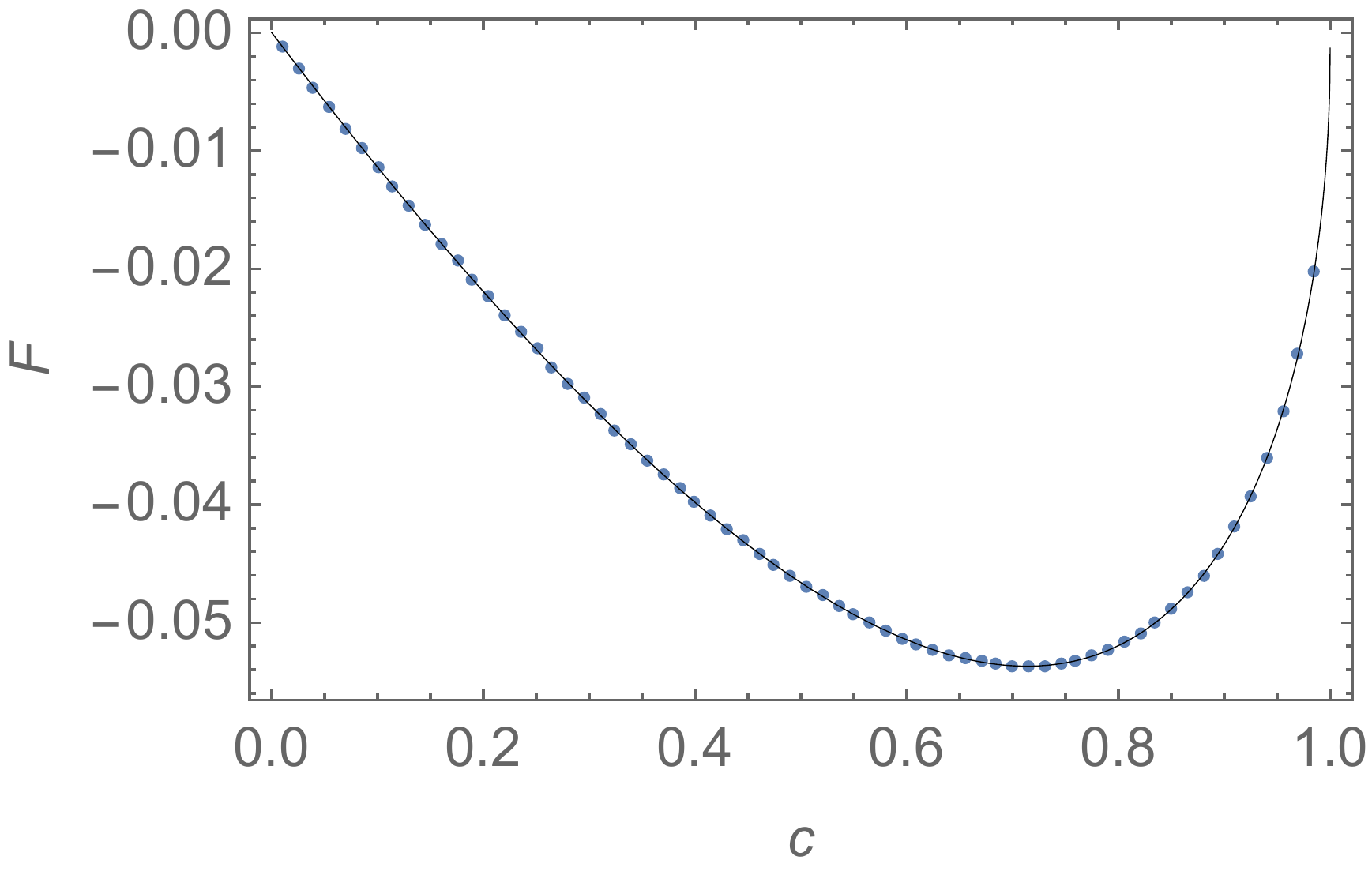}
      \caption{Four fitted constants (Eqs.(\ref{eq:fitB}) to (\ref{eq:fitF})) in the zeroth-order expansion of the BRDF as a function of single-scattering albedo $c$.  Approximation (thin) vs Exact (dots).}
      \label{fig-fitBCDF} 
    \end{figure}

    To evaluate \autoref{eq:f0} we require the constants $A, B, C, D, E, F$.  Two of these follow from simple relations~\cite{horak61}
    \begin{equation}
      A = \frac{69 c}{128}, \quad E = \frac{15}{128} (1-c) c \left(\frac{4 c}{3}+3\right).
    \end{equation}
    The other four constants involve the moments of the $H$ function and are involved equations so we fit the following approximations 
    \begin{align}
      B &= \frac{0.346689 (1-c)^{3/2}-0.777574 (1-c)+0.515357 \sqrt{1-c}-0.084463}{0.182602
      (1-c)-0.665502 \sqrt{1-c}+0.964893} \label{eq:fitB} \\
      C &= \frac{-5602.45 (1-c)^{3/2}+7487.99 (1-c)-2567.74 \sqrt{1-c}+682.848}{1480.25
      (1-c)-4008.33 \sqrt{1-c}+5850.6} \\
      D &= \frac{166.883 (1-c)^{3/2}-327.428 (1-c)+160.397 \sqrt{1-c}+0.285529}{596.423
      (1-c)-412.984 \sqrt{1-c}+674.191} \\
      F &= \frac{266.063 (1-c)^{3/2}-21.9141 (1-c)-242.16 \sqrt{1-c}-1.9209}{215.773 (1-c)+457.42
      \sqrt{1-c}+1499.9}. \label{eq:fitF}
    \end{align}
    The accuracy of these approximations is shown in \autoref{fig-fitBCDF}.  
    
    \begin{figure}
      \centering
      \includegraphics[width=\linewidth]{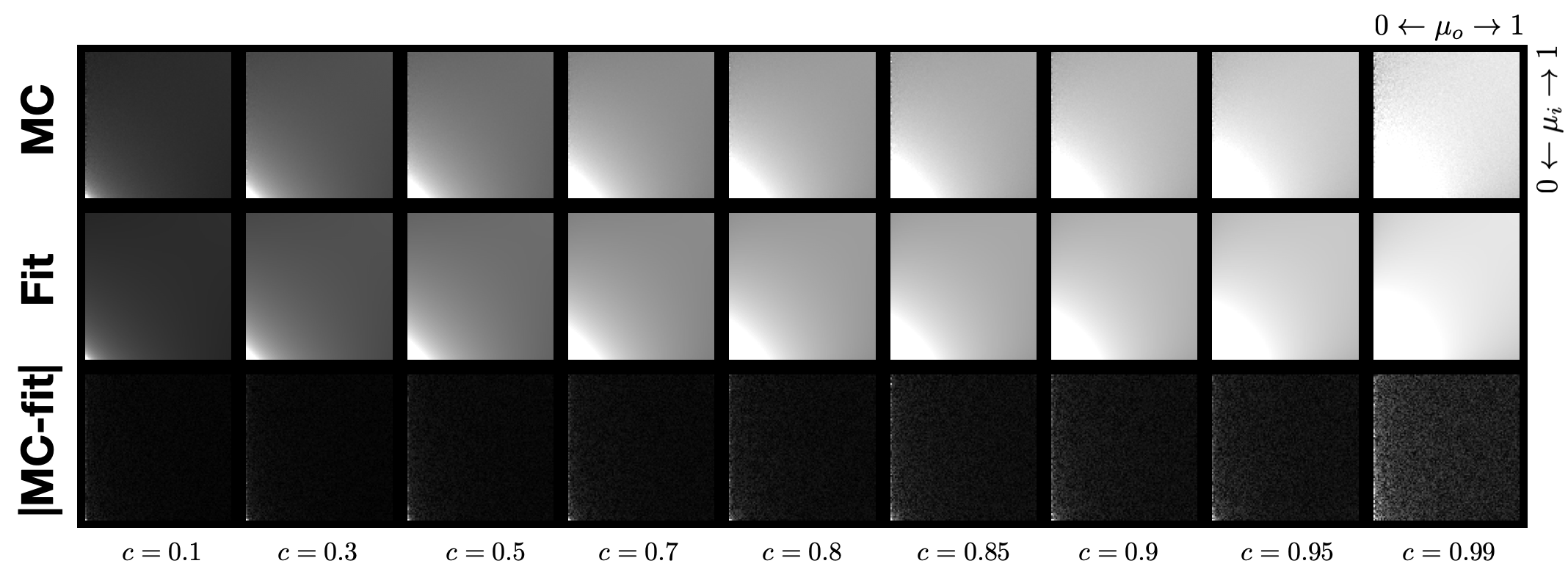}
      \caption{Error analysis of our approximate fit for the zeroth-order transfer matrix $f^{(0)}(\mu_i,\mu_o)$ versus a Monte Carlo (MC) reference simulation.}
      \label{fig-f0-MC-compare} 
    \end{figure}
    
    \subsubsection{Single-scattering}

    \autoref{eq:f0} contains all orders of scattering.  In order to use the single-scattering result exactly, we need to subtract out the approximate single-scattering from $f^{(0)}(\mu_i,\mu_o)$ using the three-term phase function appoximation,
      \begin{equation}
        f_m^{(0)}(\mu_i,\mu_o) = f^{(0)}(\mu_i,\mu_o) -\frac{c \left(45 \mu _i^2 \left(3 \mu _o^2-1\right)+256 \mu _i \mu _o-45 \mu
        _o^2+207\right)}{768 \pi  \left(\mu _i+\mu _o\right)}.
      \end{equation}

  \subsection{Albedo Mapping}
    We use the following fitted approximations for mapping between single-scattering albedo $c$ of the particles in the material and $k_d$, the spherical/bond albedo of the material (the \emph{diffuse color} $k_d$ is more intuitive for artist control),
    \begin{align}
      c &= \frac{1-1.00425 \left(1-k_d\right){}^{2.67103}}{1-0.219924
      \left(1-k_d\right){}^{2.44559}} \label{eq:kd-to-c}, \\
      k_d &= \frac{-0.453029 (1-c)-0.544162 \sqrt{1-c}+1}{1.42931 \sqrt{1-c}+1}.
    \end{align}

  \section{Fast Variant}

\begin{figure}
      \centering
      \includegraphics[width=\linewidth]{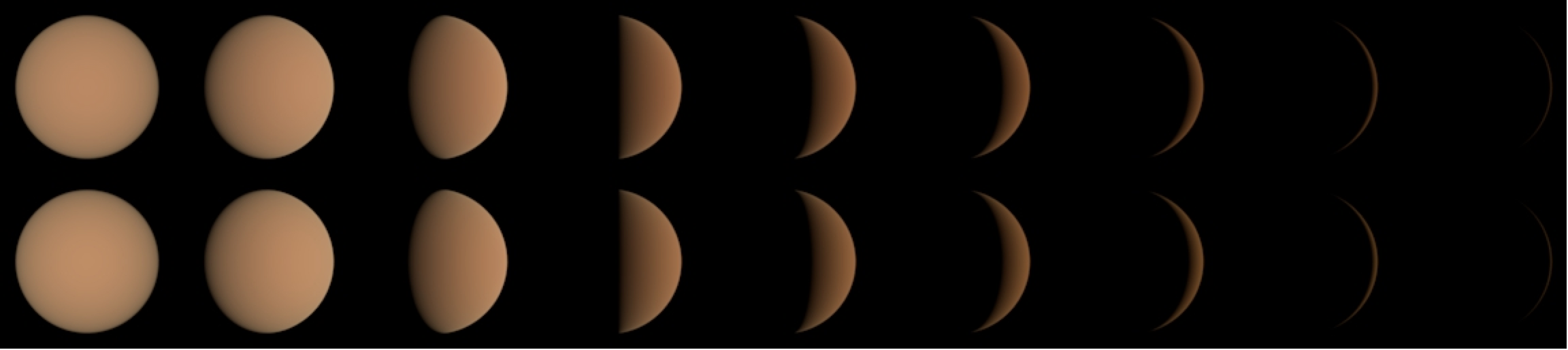}
      \caption{Comparison of our accurate analytic derivation (top) to our fast variant (bottom).}
      \label{fig-fast-nonfast-spheres} 
    \end{figure}

    The analytic fitting in the previous sections still amounts to considerable compute relative to other analytic BRDFs and maybe be too costly for real-time applications in particular.  For less accuracy and more efficiency we also found the following approximation using symbolic regression software TuringBot,
    \begin{equation}
      f_r(\dir_i,\dir_o) = \max \left(0,f_1(\dir_i,\dir_o) + \frac{0.0151829 (c-0.249978) \left(| \phi | +\sqrt{\mu _i \mu _o}\right)}{\frac{\cos
      ^{-1}(S)}{S}+0.113706}+0.234459 \text{kd}^{1.85432}\right),
    \end{equation}
    where $S=\sqrt{1-\mu _i^2} \sqrt{1-\mu _o^2}$.  \autoref{fig-fast-nonfast-spheres} compares the accuracy of this approximation to that of the previous section.

  \section{Results}

\begin{figure}
    \centering
    \includegraphics[clip, trim=0.0cm 5.0cm 0.0cm 5.0cm, width=\linewidth]{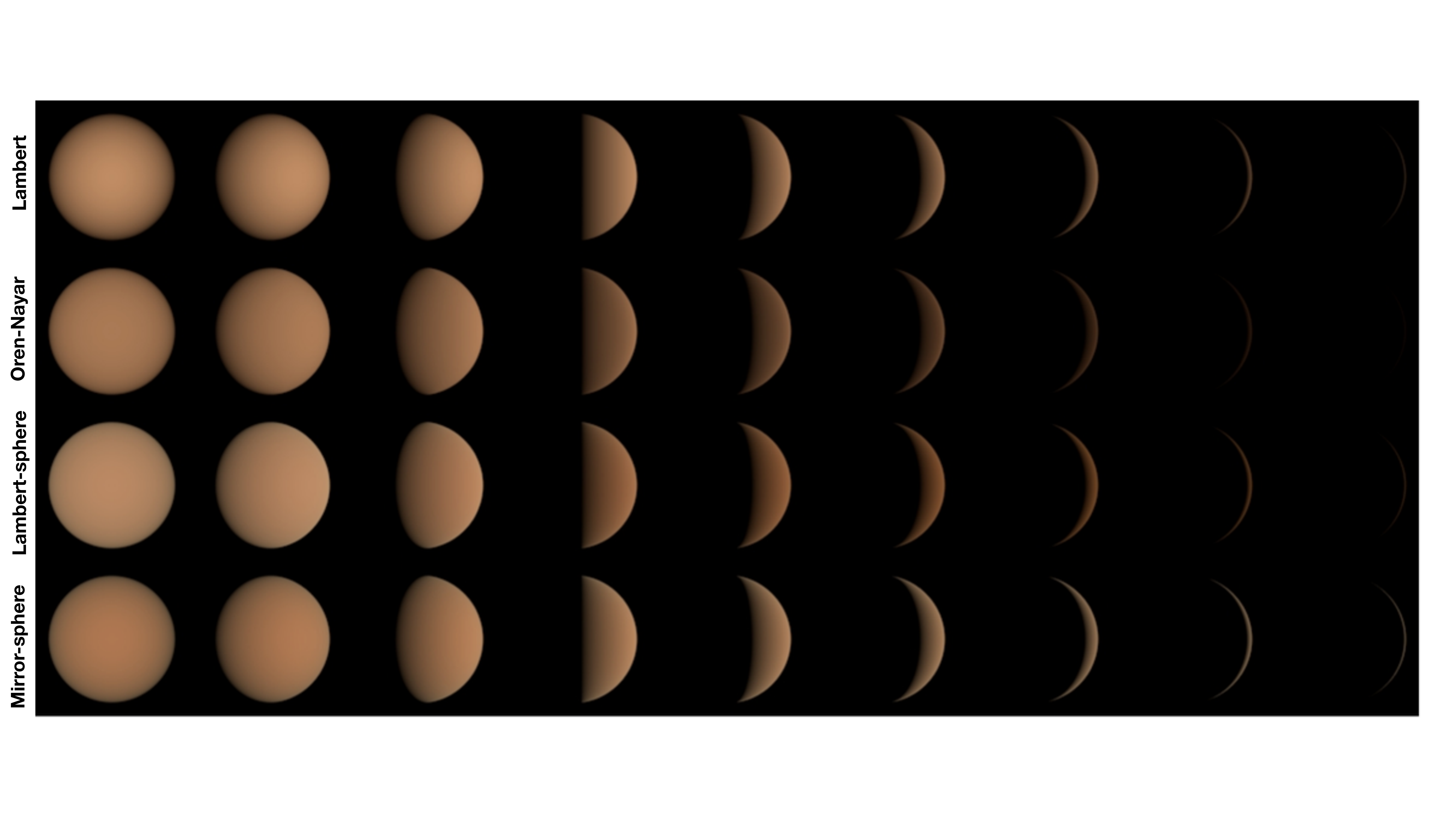}
    \caption{Comparison of our new BRDF to two height-field BRDFs and one volumetric BRDF (mirror-sphere) for a variety of lighting directions.  Note the increased back-scattering and saturated silhouettes for back-lighting compared to the other models.}
    \label{fig-dusty-compare} 
  \end{figure}

\begin{figure}
    \centering
    \subfigure[Lambertian]{\includegraphics[width=.32\linewidth]{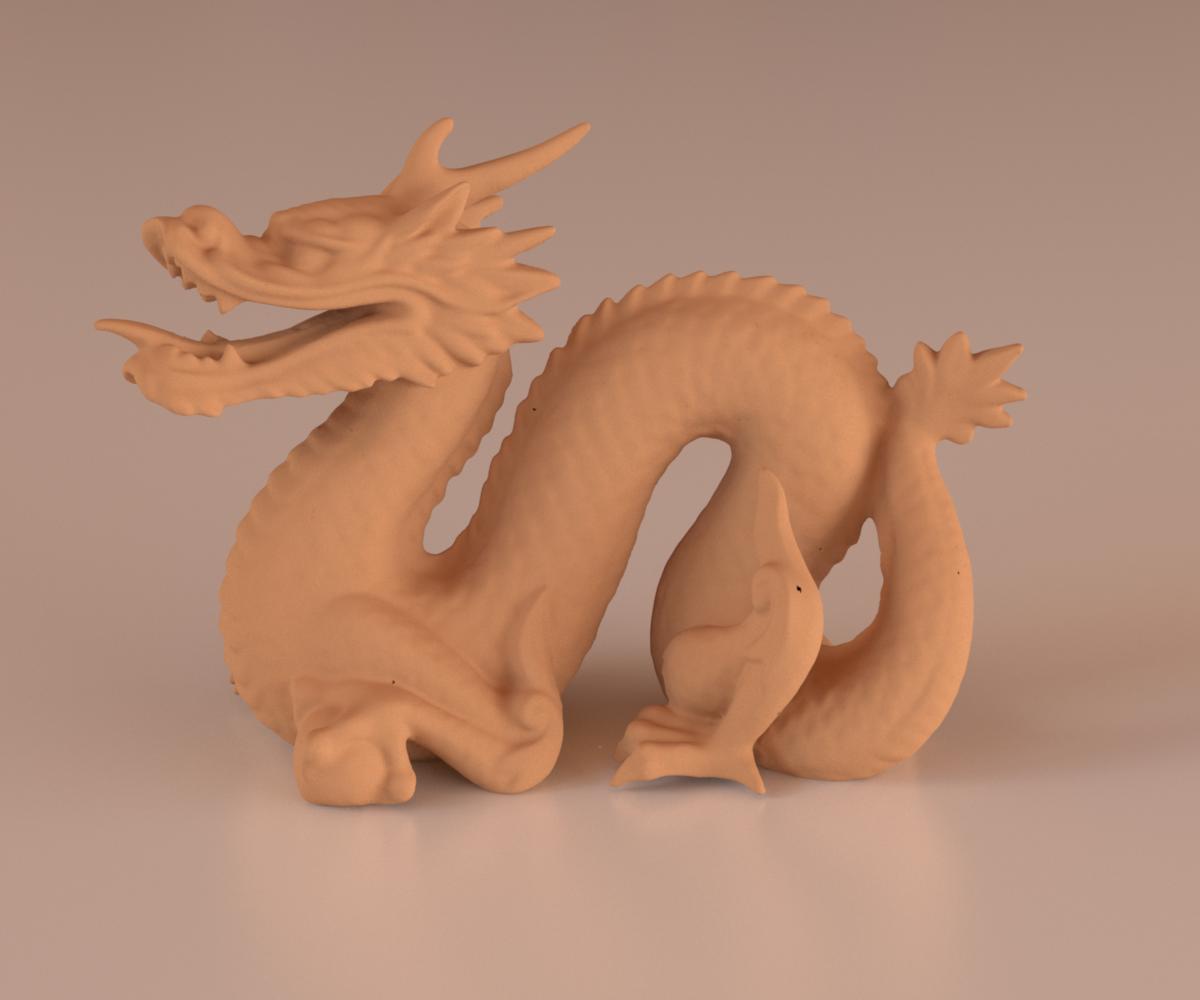}}
    \subfigure[Mirror-sphere]{\includegraphics[width=.32\linewidth]{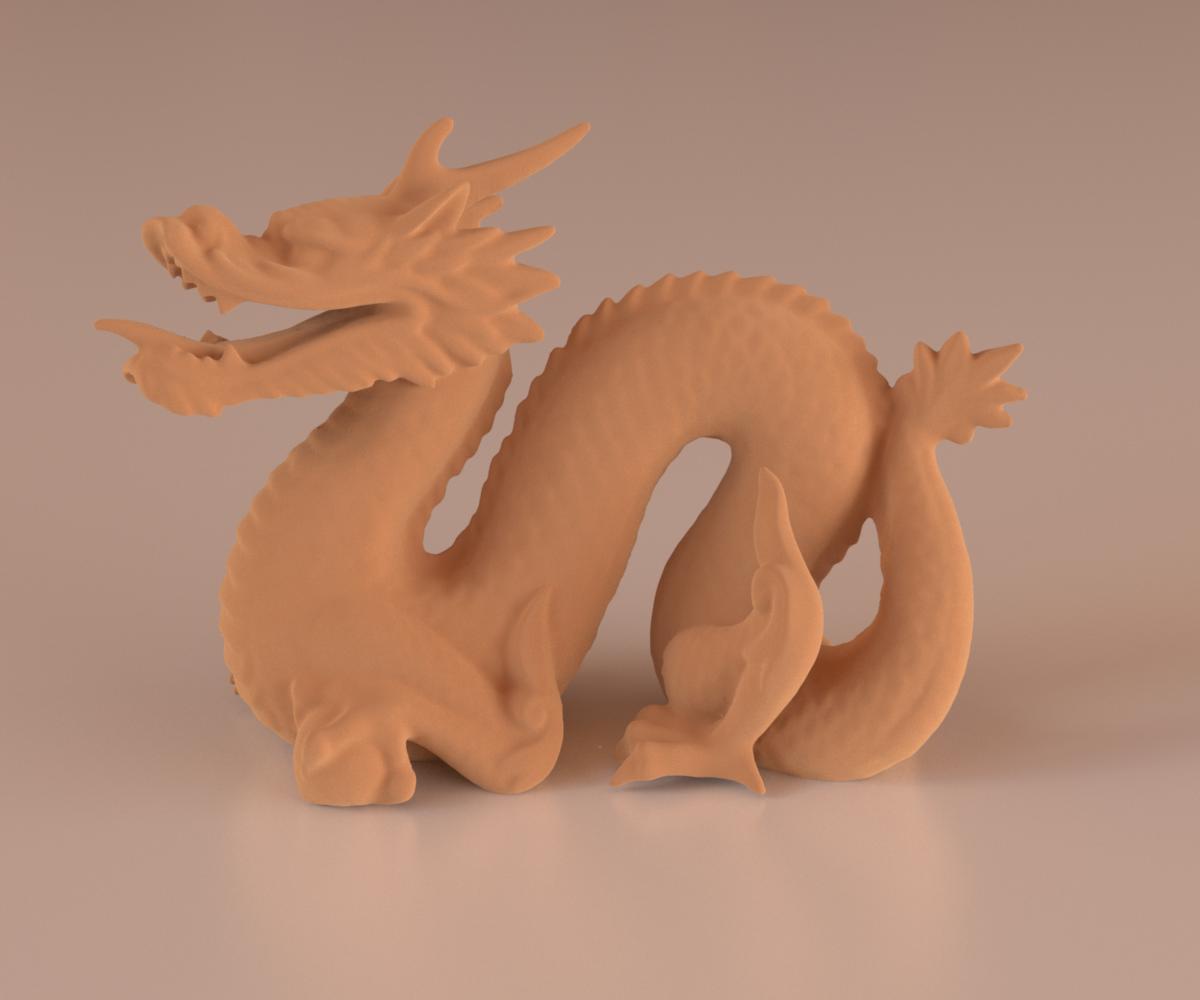}}
    \subfigure[Lambert-sphere (ours)]{\includegraphics[width=.32\linewidth]{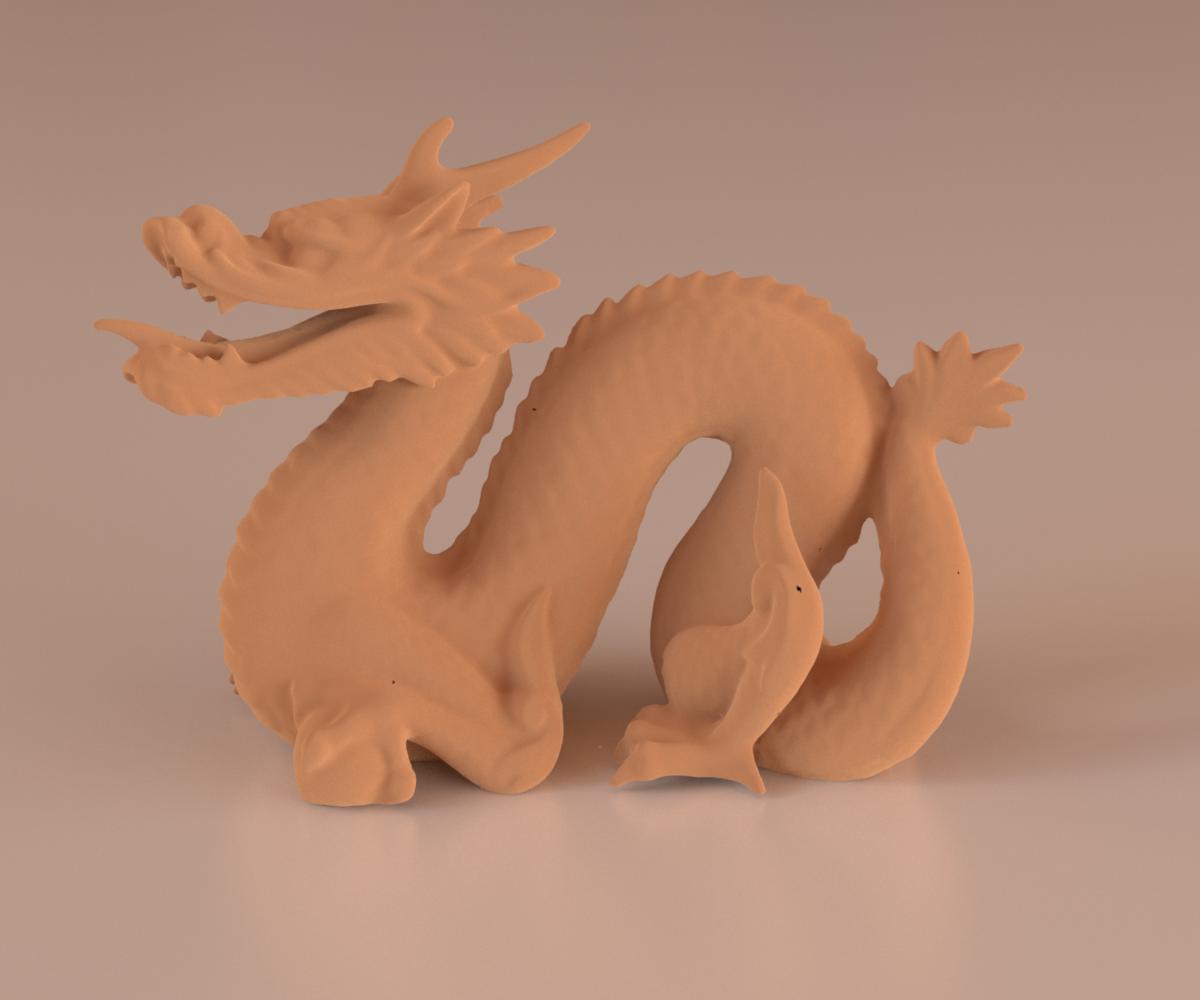}}
    \caption{Comparison of 3 diffuse BRDFs.}
    \label{fig-dragons} 
  \end{figure}

  We implemented our BRDF in Mitsuba \cite{jakob2010mitsuba}, exposing the diffuse color parameter $k_d$, which is converted to a particle albedo using \autoref{eq:kd-to-c}.  The BRDF is implemented using \autoref{eq:BRDF} and related equations.  The BRDF differs significantly in appearance from traditional diffusive BRDFs such as Lambertian, Oren-Nayar \shortcite{oren1994generalization}, and Chandrasekhar's BRDF for mirror sphere particles (\autoref{fig-dusty-compare}).  Note the increased backscattering and saturated colors for back lighting compared to the other models.  The BRDF looks most similar to the other volumetric BRDF (Chandrasekhar's), but the bright silhouettes of Chandrasekhar's are avoided with our new BRDF (\autoref{fig-dragons}).

\begin{figure}
    \centering
    \includegraphics[clip, trim=12.5cm 0.0cm 7.8cm 0.0cm, width=\linewidth]{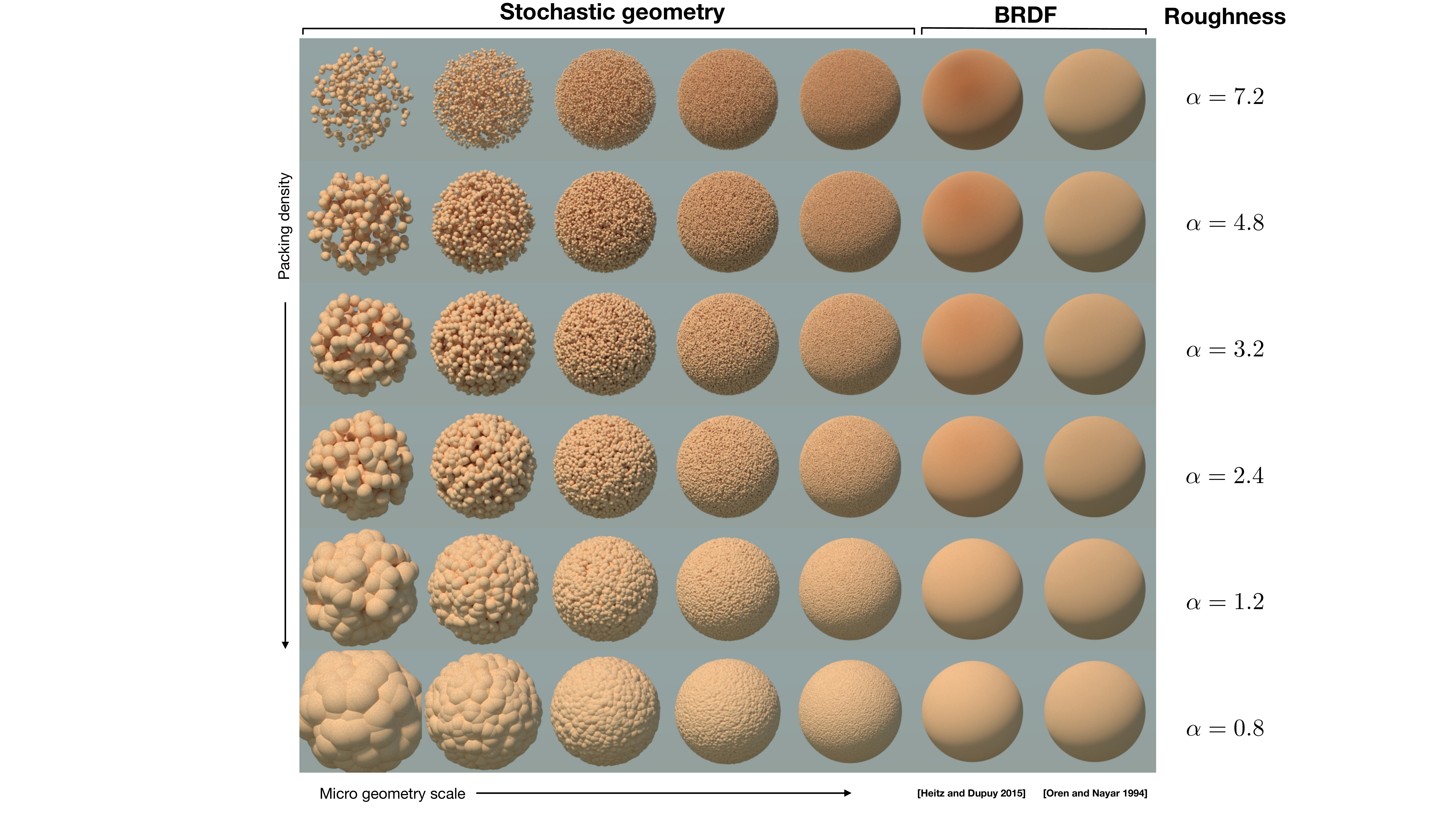}
    \caption{An array of granular microgeometries where the packing density and microgeometry scale vary while the diffuse albedo of the Lambertian geometry is fixed.  The fine-scale appearance darkens and saturates with increased sparsity because more collisions occur (on average) before a given ray escapes the sphere and finds the light source.  For dense packings (bottom rows), a random height-field is a reasonable assumption for the surface and height-field BRDFs accurately approximate the reflectance of fine-scale microgeometry (bottom right).  However, for sparse granular media (top rows), the height-field assumption is inconsistent with the microgeometry, causing either dark artifacts [Heitz and Dupuy 2015] or bright unsaturated results [Oren and Nayar 1994] (top right).}
    \label{fig-ON-grid} 
  \end{figure}

  Our volumetric BRDF can model the appearance of sparse granular materials that height-field models cannot.  \autoref{fig-teaser} shows how our model closely matches the granular microgeometry of sparse Lambertian spheres.  To compare to various sphere packings that violate the assumptions of classical radiative transfer (that the scatterers are spatially independent), in \autoref{fig-ON-grid} we consider a denser array of packings with the Lambertian albedo held fixed in all cases.  Height-field models can do a reasonable job at approximating a sphere cluster geometry up until about roughness $\alpha = 3$, although this requires significant stochastic evaluation to account for the many orders of scattering required to represent the full BRDF.  (The Oren Nayar model only accounts for 2 bounces, and does not extend to such a range of roughnesses).   Around this point the height-field assumption is inconsistent and a spherical NDF would be more appropriate \cite{dupuy2016additional}.  Past this roughness level, the rough diffuse Beckmann BRDF \cite{heitz2015implementing} shows dark artifacts because the roughness simply scales a height field until the profile is unreasonably spiky \cite{dupuy2016additional}.

  \section{Conclusion}

  We have presented two new importance-sampling schemes for the Lambertian-sphere phase function and have used a three-term expansion of this phase function to derive an analytic BRDF for dusty / granular media.  This BRDF differs significantly from other analytic diffuse BRDFs and may find application in a number of areas.  Future work includes investigation of measured data to see if this behaviour is found in real-world materials.  We also want to derive a single analytic BRDF that blends from Lambertian through height-field models and into our new BRDF by considering a sequence of microgeometry like that illustrated in \autoref{fig-ON-grid} modeled using a spherical Gaussian NDF and a novel form of Smith's model for handling such media \cite{deon16anisotropic}.  In this way the classical half space with spatially-uncorrelated spherical particles becomes the \emph{infinite roughness} endpoint of a new expressive family of rough diffuse materials.

\bibliographystyle{acmsiggraph}
\bibliography{lambertsphereBRDF}

\end{document}